\documentstyle[12pt]{article}
              
\title{Localization of Events in Space-Time.}
\author{M. Toller   \\ 
Dipartimento di Fisica dell'Universit\`a, Trento  \\
I.N.F.N. gruppo collegato di Trento, Italia}
 
\newtheorem{proposition}{Proposition}
\newtheorem{definition}{Definition}
\begin{document} 
\maketitle                             
                 
\begin{abstract}
The present paper deals with the quantum coordinates of an event in space-time, individuated by a quantum object. It is known that these observables cannot be described by self-adjoint operators or by the corresponding spectral projection-valued measure. We describe them by means of a positive-operator-valued (POV) measure in the Minkowski space-time, satisfying a suitable covariance condition with respect to the Poincar\'e group.  This POV measure determines the probability that a measurement of the coordinates of the event gives results belonging to a given set in space-time. We show that this measure must vanish on the vacuum and the one-particle states, which cannot define any event.  We give a general expression for the Poincar\'e  covariant POV measures.  We define the baricentric events, which lie on the world-line of the centre-of-mass, and we find a simple expression for the average values of their coordinates.  Finally, we discuss the conditions which permit the determination of the coordinates with an arbitrary accuracy.
\bigskip  \bigskip

\end{abstract}
\newpage

\section{Introduction.}   

The aim of the present article is to study how a physical quantum system can define (with some indetermination) a point in Minkowski space-time, namely an event. This study, even if it has a rather formal character, may help to clarify the operational meaning of the concept of event, namely its definition in terms of observables. Some results about this problem have been given in \cite{Toller}. The most natural approach is to consider the space-time coordinates of the event as quantum observables described by the hermitian operators $X^{\alpha}$, ${\alpha = 0, 1, 2, 3}$. Operators of this kind have been defined in ref.\ \cite{JR} in the case of a relativistic system of zero mass particles, which has a symmetry under dilatations.

If we indicate by $P^{\alpha}$ the self-adjoint operators that describe the components of four-momentum, it is natural to assume that, in a suitable dense domain of the Hilbert space $\cal H$, we have ($\hbar = c = 1$, $g^{00}=1$)
\begin{equation} \label{Commu}
[P^{\alpha}, X^{\beta}] = i g^{\alpha \beta}, 
\end{equation}  
or, in the translation invariant   domain where the operators $X^{\alpha}$ are defined, 
\begin{equation} \label{Commu2}
\exp(-ix_{\alpha} P^{\alpha}) X^{\beta} \exp(ix_{\alpha} P^{\alpha}) =  X^{\beta} +x^{\beta}. 
\end{equation}
If the operator $p_{\alpha} X^{\alpha}$ is self-adjoint, we have
\begin{equation}
\exp(ip_{\alpha} X^{\alpha}) P^{\beta} \exp(-ip_{\alpha} X^{\alpha}) =  P^{\beta} +p^{\beta} 
\end{equation} 
and it follows that the joint spectrum of the four-momentum operators $P^{\alpha}$ is invariant under translations in the direction of the four-vector $p^{\alpha}$. Since this joint spectrum is contained in the future cone, $p_{\alpha} X^{\alpha}$ cannot be self-adjoint. It follows that the operators  $X^{\alpha}$ cannot have a spectral representation and the statistical interpretation of the corresponding observables requires some particular attention.
                      
The argument given above, discussed by Wightman \cite{Wightman}, is an immediate generalization of a well known argument due to Pauli \cite{Pauli} concerning the time observable $T$, namely the quantity obtained by reading a quantum clock. It satisfies the commutation relation
\begin{equation} 
{ d T \over dt} =  i [H, T] = 1, 
\end{equation} 
where $t$ is the usual time parameter, measured by a classical external clock. If $T$ is self-adjoint this equation contradicts the fact that the spectrum of the Hamiltonian $H$ is bounded from below. 

Our coordinate $X^0$ is strictly related to the reading of a clock, but it is more similar to a time-of-arrival observable \cite{Allcock, MBM, GRT, Leon, DM, Delgado, AOPRU, MSP}, namely the time registered by a classical clock when some event happens, for instance a quantum particle reaches a given point, or two quantum particles collide. If we consider a quantum clock, the time-independent observable
\begin{equation}
X^0 = t - T
\end{equation}
is the time t measured by a classical clock when the quantum clock gives  $T = 0$, and it is a typical time-of-arrival observable.  Its commutator with the Hamiltonian $H = P^0$ is given by eq.\ (\ref{Commu}). 

Here we deal with an ``indirect'' measurement of a time-of-arrival, namely the measurement operation can be performed at any time $t$ and we use the equations of motion, which are supposed to be known. A different and more difficult problem is the ``direct'' measurement of a time-of-arrival, performed by means of operations lasting a long time and detecting immediately the event at the time at which it happens.
 
The quantum time problem has been discussed by several authors, see for example \cite{Armstrong, SaW, EF, AB, Rosenbaum, ORG, Peres, UW, BJ, Mayburov}, besides the ones cited above.  A satisfactory solution is obtained \cite{Holevo, BGL, Giannitrapani, Giannitrapani1} by writing a generalized spectral representation 
\begin{equation} 
T = \int t \, d\tau(t), 
\end{equation}
where $\tau$ is a normalized positive-operator-valued (POV) measure on the real line. Since T is not self-adjoint, $\tau$ cannot be a projection-valued measure (for a different point of view, see \cite{GRT}). The POV measure $\tau$ is not uniquely determined by the operator $T$, but it describes the time observable completely, since the probability that the result of a time measurement is contained in an interval $I$ is given by
\begin{equation} 
{\cal P}(I) = (\psi , \tau(I) \, \psi), 
\end{equation} 
\begin{equation} 
\tau(I) =  \int_I d\tau(t), 
\end{equation} 
where the normalized vector $\psi$ describes the quantum state of the clock.

Observables of this kind have been considered for different purposes by several authors \cite{Holevo, Davies, BLM, BGL2, BGL3}.  The operator $\tau(I)$ represents a test (or an effect) \cite{Ludwig, Ludwig2, Giles}, namely a mixed yes-no observable. If we decompose the real line into a set of non-overlapping intervals $I_1,\ldots, I_n$, the operators $\tau(I_1),\ldots, \tau(I_n)$ represent a multi-bin test. One can show \cite{Lubkin, Lubkin2} that for any multi-bin test one can find a corresponding measuring instrument, if there are no limitations to the choice of the interaction Hamiltonian. This result legitimates the use of observables defined by POV measures within the standard formalism of quantum theory.

The aim of the present paper is to apply the POV measure formalism to the four space-time coordinates $X^{\alpha}$ of an event  measured with respect to a classical reference frame.  The quantities $X^1, X^2, X^3$ should not be confused with the self-adjoint Newton-Wigner coordinates of a particle \cite{Wightman, NW, CM, Mourad}, which do not commute with the Hamiltonian $P^0$, since the position of the particle changes with time. The coordinates of an event are clearly time-independent.
                   
A particular need of a clear treatment of the quantum properties of the space-time coordinates arises when one considers the limitations to the measurements of time and length, which appear when one tries to merge quantum theory, relativity and gravitation \cite{Mead, Ferretti, NVD, Garay, DFR, Gibbs, KMN, Camelia}. From this point of view, our treatment in the absence of gravitation is just a preliminary but necessary exercise. In fact, we have to remember that in general relativity the physical meaning of the coordinates is a delicate problem even in the absence of quantum effects \cite{Rovelli2, BM}.

In Section 2 we describe the POV measures in the Minkowski space-time which are covariant with respect to the space-time translations. In Section 3 we impose the Poincar\'e covariance condition and we give an explicit general formula for these POV measures. We also discuss the constraint which appears in the presence of a symmetry under dilatations. We shall not consider in this article the conditions imposed by the covariance under space and time reflections, when the theory considered has these symmetries.

It is expected that, given a suitable physical object, the choice of the POV measure is not uniquely determined. In fact there is a large arbitrariness in the choice of the conventions which define the event in terms of the properties and the motion of the object. It follows that it is interesting to study more restricted classes of POV measures obtained by imposing some further constraints.  In Section 4 we discuss the ``baricentric''  events, which lie, exactly or approximately, on the world-line of the centre-of-mass of the object that defines them. In Section 5 we give explicit expressions for the operators $X^{\alpha}$ and we compare our results with the ones of ref.\ \cite{JR}. In Section 6 we study the conditions which permit the determination of the coordinates of an event with an arbitrary accuracy.

\section{Translation covariant POV measures.}  

Following the ideas introduced above, we consider a POV measure $\tau(I)$ on the Minkowski space-time $\cal M$. If the normalized vector $\psi \in {\cal H}$ describes, in the Heisenberg picture, the state of the system that defines the event, the probability that the event is found in the Borel set $I \subset {\cal M}$ is given by
\begin{equation} 
{\cal P}(I) = 
(\psi, \tau(I) \psi). 
\end{equation}
It is necessary to make clear that we are dealing with ``indirect'' measurements of the coordinates of an event, namely the test $\tau(I)$ is not measured by means of physical operations performed in the space-time region $I$. For this reason we do not require that the operators $\tau(I)$ and $\tau(I')$ commute if the regions $I$ and $I'$ are space-like separated. Actually, it has been shown \cite{Giannitrapani2} that $\tau(I)$ cannot be a quasi local observable \cite{HK}. The normalization condition 
\begin{equation} \label{Normal} 
\tau({\cal M}) = 1
\end{equation} 
means that an event is certainly detected at some point of space-time. We shall show that, in general, this is not true for an arbitrary choice of the state $\psi$; for instance the vacuum state cannot define any event. Therefore we adopt the weaker assumption  
\begin{equation}  \label{Bound}
0 < \tau({\cal M}) \leq 1.
\end{equation} 
Then we put  
\begin{equation} \label{Coordinates}
X^{\alpha} = \int_{\cal M} x^{\alpha} \, d\tau(x). 
\end{equation}
Since these operators cannot be self-adjoint, $\tau$ cannot be a projection-valued measure. 

We indicate by $\tilde{\cal P}$ the universal covering of the proper orthochronous Poincar\'e group $\cal P$. For its elements we use the notation $(x, a)$, where $x$ is a four-vector which describes a translation  and $a \in SL(2, C)$. $\Lambda(a)$ is the $4 \times 4$ Lorentz matrix corresponding to $a$. If $U(x, a)$ is the unitary representation of $\tilde{\cal P}$ that acts on the space $\cal H$, we require that 
\begin{equation}  \label{Covariance}
U^{\dagger}(x, a) \tau(\Lambda(a)I + x) U(x, a) = \tau(I).
\end{equation} 
This means that the POV measure $\tau$ and the representation $U$ of $\tilde{\cal P}$ form a ``system of covariance''\cite{CH}. If $\tau$ were a projection-valued measure, we should have a ``system of imprimitivity'' \cite{Mackey}. Of course, the covariance assumption is valid if no external objects intervene in the definition of the event.    

It is clear that the covariance and the boundedness conditions (\ref{Bound}) do not determine the POV measure $\tau$ uniquely. For instance, if $K$ is an unitary operator that commutes with all the operators $U(x, a)$, the POV measure
\begin{equation} 
\tau'(I) = K^{\dagger}\tau(I)K
\end{equation}
satisfies the required conditions as well as $\tau$.  The covariance condition (\ref{Covariance}) is satisfied even if $K$ is not unitary. 

In the rest of the present Section we consider a $d$-dimensional space-time and we use only the covariance with respect to the space-time translation group, that can be written in the form
\begin{equation} \label{Covariance2}
\exp(-ix_{\alpha} P^{\alpha}) \tau(I + x) \exp(ix_{\alpha} P^{\alpha}) = \tau(I).
\end{equation} 
From this equation and eq.\ (\ref{Coordinates}) we obtain
\begin{equation}  \label{Commu3}
\exp(-ix_{\alpha} P^{\alpha}) X^{\beta} \exp(ix_{\alpha} P^{\alpha}) =  X^{\beta} +x^{\beta} \tau({\cal M}), 
\end{equation} 
which coincides with eq.\ (\ref{Commu2}) if the measure $\tau$ is normalized. 

If we also assume that the momentum spectrum is contained in the closed future cone $\overline V$, as a consequence of eq.\ (\ref{Covariance2}) we obtain the following result: 
\begin{proposition} \label{Positivity}
If $I \subset {\cal M}$ is a non-empty open set, we have  
\begin{equation} 
\tau(I) > 0.
\end{equation}
Moreover, the equality 
\begin{equation} 
(\psi, \tau(I) \psi) = 0
\end{equation}   
implies that
\begin{equation} 
(\psi, \tau({\cal M}) \psi) = 0.
\end{equation}  
\end{proposition} 

If $\tau(I) = 0$, we also have  $\tau(I + x) = 0$ for any choice of the vector $x$ and therefore  $\tau({\cal M}) = 0$ in contradiction with eq. (\ref{Bound}). In order to prove the second part of the Proposition, we consider two open sets  $I'$ and $I''$  with the property  
\begin{equation} 
I' + I'' \subset I.
\end{equation}
Then we have
\begin{displaymath}
(\psi, \tau(I' + x) \psi) =
(\exp(-ix_{\alpha} P^{\alpha}) \psi,  \tau(I') \exp(-ix_{\alpha} P^{\alpha}) \psi) = 0 
\end{displaymath}
\begin{equation} 
{\rm for} \quad x \in I'',
\end{equation}   
namely
\begin{equation} 
(\tau(I'))^{1/2} \exp(-ix_{\alpha} P^{\alpha}) \psi = 0 
\qquad {\rm for} \quad x \in I''.
\end{equation}  
This expression is the limit of a vector-valued function analytic in the tube defined by ${\rm Im}\, x \in -V$, where $V$ is the open future cone. An application of the edge-of-the-wedge theorem \cite{StW} shows that this analytic function vanishes in the whole tube and it follows that 
\begin{equation} 
(\psi, \tau(I' + x) \psi) = 0 
\end{equation}
for any real value of $x$. The announced result follows from the additivity property of the measure.

From Proposition \ref{Positivity}, we obtain another proof that $\tau$ cannot be a projection-valued measure. In fact, if $I$ has a non-empty interior, $\tau(I)$ cannot be a projection operator different from $\tau({\cal M})$.  We also see that the localization of an event in a bounded region $I$ cannot be considered as a ``property'' of the system. 

The problem of finding a general representation for a covariant POV measure has been studied by several authors \cite{CH, Davies2, Scutaru, Holevo2, Kholevo}. Here we give, for easier reference, a self-contained treatment of the particular case in which we are interested. In the meantime, we introduce the notations necessary for further developments.

The Hilbert space $\cal H$ of the theory is given by the direct integral \cite{Mackey, Dixmier}
\begin{equation} 
{\cal H} = \int^{\oplus} {\cal H}(k) \, d\mu(k),
\end{equation}    
where $\mu$ is a measure in the $d$-dimensional momentum space.  If we choose suitable bases in the spaces ${\cal H}(k)$, its elements are described by the wave functions $\psi_{\sigma}(k)$, defined in regions of momentum space that can depend on $\sigma$. We adopt the convention that  the wave functions vanish outside the region in which they are defined. The norm is given by
\begin{equation} \label{Norm}
\|\psi\|^2 = 
\int \sum_{\sigma} |\psi_{\sigma}(k)|^2  \, d\mu(k).  
\end{equation}  

The measure $\mu$ can be decomposed \cite{Halmos} into a part $\mu'$ absolutely continuous with respect to the Lebesgue measure and a part $\mu''$ which has a support with vanishing Lebesgue measure. The Hilbert space has the corresponding decomposition
\begin{equation} 
{\cal H} = {\cal H}' \oplus {\cal H}''.
\end{equation}
We can rescale the wave functions (outside the support of $\mu''$) and replace the measure $d \mu'$ by the equivalent measure $f_S(k) \, d^d k$, where $f_S(k)$ is the characteristic function of the support $S$ of $\mu'$. With our convention on the wave functions, this factor can be omitted. If we indicate by $P'$ the projection operator on the subspace ${\cal H}'$, we have
\begin{equation}  \label{Projection}
(\psi, P' \psi) = \int \sum_{\sigma} |\psi_{\sigma}(k)|^2 \, d^d k.
\end{equation} 

We consider the dense translation invariant linear space ${\cal D} \subset {\cal H}$ composed of the wave functions in momentum space which are infinitely differentiable, fast decreasing and not vanishing only for a finite set of values of the index $\sigma$. They have the property
\begin{equation}
\|\psi\|_r^2 = 
\int \sum_{\sigma} |(1+k_0)^r \psi_{\sigma}(k)|^2  \, d\mu(k) < \infty,
\qquad r = 0, 1,\ldots, 
\end{equation}
which implies that $F(P) \psi \in {\cal H}$ for any choice of the polynomial function  $F(P)$ of the momentum operators. We define the topology of $\cal D$ by means of this family of norms.

 The convolution of the numerical measure $(\psi, \tau(x) \phi)$ with the function $g(x)$, continuous and with compact support, is a function of $x$ given by 
\begin{displaymath}
\left(\psi, \int g(x - y) \, d \tau(y) \, \phi \right) = 
\end{displaymath}
\begin{equation} 
= \left(\exp(-ix_{\alpha} P^{\alpha}) \psi, \int g(-y) \, d \tau(y) \, \exp(-ix_{\alpha} P^{\alpha}) \phi \right).
\end{equation}
Its partial derivatives are given by sums of similar expressions in which the vectors $\psi$ and $\phi$ are replaced by vectors of the kind  $F(P) \psi$ and $F'(P) \phi$, where $F(P)$ and $F'(P)$ are polynomials. If $\psi, \phi \in {\cal D}$, we see that the convolution defined above is infinitely differentiable for any choice of the continuous function $g$.  A general theorem concerning distributions \cite{Schwartz} permits one to draw the following conclusion:
\begin{proposition} 
If $\psi, \phi \in {\cal D}$, we can write  
\begin{equation}  
(\psi, \tau(I) \phi) = \int_I \rho(\psi, \phi, x) \, d^d x, 
\end{equation} 
where $\rho(\psi, \phi, x)$ in an infinitely differentiable function of $x$. In particular
\begin{equation} \label{Integral2} 
(\psi, \tau(I) \psi) = \int_I \rho(\psi, x) \, d^d x, 
\end{equation}
where
\begin{equation}  
\rho(\psi, x) = \rho(\psi, \psi, x) \geq 0.
\end{equation}
\end{proposition}  

If we introduce the set
\begin{equation}  
I(x) =\{y \in {\cal M}:  y^0 < x^0, y^1 < x^1,\ldots, y^{d-1} < x^{d-1} \}
\end{equation} 
we have
\begin{displaymath}
 \rho(\psi, \phi, x) = {\partial^d \over \partial x^0 \partial x^1 \cdots \partial x^{d-1}}  (\psi, \tau(I(x)) \phi)  =
\end{displaymath}
\begin{equation}  
= {\partial^d \over \partial x^0 \partial x^1 \cdots \partial x^{d-1}} \left(\exp(-ix_{\alpha} P^{\alpha}) \psi, \tau(I(0)) \exp(-ix_{\alpha} P^{\alpha}) \phi \right).
\end{equation} 
A simple calculation gives
\begin{equation}  
| \rho(\psi, \phi, x) | \leq  2^d \|\psi\|_d \|\phi\|_d
\end{equation}   
and we see that $\rho(\psi, \phi, x)$ for fixed values of $x$ is a continuous function of $\psi, \phi \in {\cal D}$.

From eq. (\ref{Bound}) and (\ref{Covariance2}) we obtain
\begin{equation} \label{Bound2}
\int \rho(\psi, x) \, d^d x \leq  \| \psi \|^2, 
\end{equation} 
\begin{equation}  \label{Covariance3}
\rho(\exp(iy_{\alpha} P^{\alpha}) \psi, \exp(iy_{\alpha} P^{\alpha}) \phi, x + y) = 
\rho(\psi, \phi, x).
\end{equation} 
Since $\rho(\psi, \phi, x)$ is a continuous function of $x$, it has a well defined value at $x = 0$. If $\rho(\psi, \phi, 0)$ is given, we put 
\begin{equation} \label{Rhox}
\rho(\psi, \phi, x) = \rho(\exp(-ix_{\alpha} P^{\alpha}) \psi, \exp(-ix_{\alpha} P^{\alpha}) \phi, 0)
\end{equation}                       
and the covariance condition (\ref{Covariance3}) is satisfied.

Note that $\rho(\psi, \phi, 0)$ is a continuous sesquilinear form on the space $\cal D$.  It defines a scalar product in the quotient space ${\cal D}/{\cal D}_0$ and on its completion $\tilde{\cal H}$, which is a Hilbert space.  We have indicated by  ${\cal D}_0$ the subspace of ${\cal D}$ defined by the condition  $\rho(\psi, 0) = 0$. This construction also defines a linear operator $h: {\cal D} \to \tilde{\cal H}$ and we have
\begin{equation} \label{ScalProd}
\rho(\psi, \phi, 0) = (h\psi, h\phi),
\end{equation} 
where at the right hand side there is the scalar product of the Hilbert space $\tilde{\cal H}$. 
The operator $h$ is continuous, since we have 
\begin{equation}  \label{Cont}
\| h \psi \| \leq 2^{d/2} \| \psi\ \|_d.
\end{equation}
               
We introduce a basis in the space $\tilde{\cal H}$  and we represent its element $\Psi$ by means of its components $\Psi_{\gamma}$. The norm is given by
\begin{equation} \label{Norm2}
\|\Psi\|^2 = \sum_{\gamma} |\Psi_{\gamma}|^2.
\end{equation}
From eq.\ (\ref{Cont}), using the Riesz theorem, we see that we can write
\begin{equation}  \label{Psi}
\Psi_{\gamma} = [h \psi]_{\gamma} = (2 \pi)^{- d/2} \int \sum_{\sigma} K_{\gamma \sigma}(k) \psi_{\sigma}(k) \, d \mu(k),
\end{equation}                          
where the functions $K_{\gamma \sigma}(k)$ are locally square integrable with respect to the measure $\mu$.

From eqs.\ (\ref{Rhox}) and (\ref{ScalProd}) we have 
\begin{equation} \label{Density}
\rho(\psi, x) =  \sum_{\gamma} |\Psi_{\gamma}(x)|^2,
\end{equation} 
where 
\begin{equation} \label{PsiX}
\Psi(x) = h \exp(-i x_{\alpha} P^{\alpha}) \psi,
\end{equation}
namely 
\begin{equation} \label{PsiX2}
\Psi_{\gamma}(x) = (2 \pi)^{- d/2} \int \sum_{\sigma} K_{\gamma \sigma}(k) \exp(-ix_{\alpha} k^{\alpha}) \psi_{\sigma}(k) \, d\mu(k).
\end{equation}    

From eq. (\ref{Bound2}) and  (\ref{Density}), we obtain
\begin{equation} \label{Bound3}
\int \sum_{\gamma} |\Psi_{\gamma}(x)|^2 \, d^d x \leq \|\psi\|^2
\end{equation} 
and we see that the functions $\Psi_{\gamma}(x)$ are square integrable. We also see that eq.\ (\ref{PsiX}) defines a bounded linear mapping $\hat h: {\cal D} \to \tilde{\cal H} \otimes  L_2({\cal M})$, which can be extended by continuity to the whole space $\cal H$. It follows that for all the vectors $\psi \in {\cal H}$ the POV measure $\tau$ can be defined by eq.\ (\ref{Integral2}), where the probability density $\rho(\psi, x)$ is an integrable function (in general not continuous). 

The square integrable functions $\Psi_{\gamma}(x)$ can be represented  as Fourier transforms of square integrable functions $\tilde\Psi_{\gamma}(k)$ in momentum space, namely we have
\begin{equation} \label{Fourier}
\Psi_{\gamma}(x) = (2 \pi)^{- d/2} \int \tilde\Psi_{\gamma}(k) \exp(-ix_{\alpha} k^{\alpha}) \, d^d k.
\end{equation}
By comparing  eqs.\  (\ref{PsiX2}) and  (\ref{Fourier})  we obtain the following equality between measures
\begin{equation} 
\sum_{\sigma} K_{\gamma \sigma}(k) \psi_{\sigma}(k) \, d\mu(k) = \tilde\Psi_{\gamma}(k) \, d^d k.
\end{equation}       
Since this measure vanishes on the sets of zero Lebesgue measure, we can drop the singular part $\mu''$ of $\mu$ and keep only the absolutely continuous part. Then we can write
\begin{equation} \label{PsiX3}
\Psi_{\gamma}(x) = (2 \pi)^{- d/2} \int \sum_{\sigma} K_{\gamma \sigma}(k) \exp(-ix_{\alpha} k^{\alpha}) \psi_{\sigma}(k) \, d^d k.
\end{equation} 

By substitution into eq.\ (\ref{Bound3}) we obtain the condition
\begin{equation}
\int \sum_{\gamma} \left|\sum_{\sigma} K_{\gamma \sigma}(k) \psi_{\sigma}(k) \right|^2 \, d^d k \leq \|\psi\|^2,
\end{equation} 
which is equivalento to the following condition valid for almost all the values of $k \in S$: 
\begin{equation} \label{Condition}
\sum_{\gamma} \left|\sum_{\sigma} K_{\gamma \sigma}(k) c_{\sigma} \right|^2  \leq \sum_{\sigma} |c_{\sigma}|^2.
\end{equation}    
This formula means that the matrices $K_{\gamma \sigma}(k)$ represent bounded operators $K(k): {\cal H}(k) \to \tilde{\cal H}$ with $\|K(k)\| \leq 1$.  

If the subspace ${\cal H}''$ is not reduced to zero, the normalization condition (\ref{Normal}) cannot be satisfied, but we can consider the weaker condition
\begin{equation} \label{Normal1} 
\tau({\cal M}) = P',
\end{equation} 
where $P'$ is the projection operator on the subspace ${\cal H}'$ defined by eq.\ (\ref{Projection}).  If we impose this condition we obtain
\begin{equation} \label{Normal2}
\sum_{\gamma} \overline{K_{\gamma \sigma}(k)} K_{\gamma \sigma'}(k) = \delta_{\sigma \sigma'}.
\end{equation} 
This equation means that the operators $K(k)$ are isometric.  

In conclusion we have:                                                                       
\begin{proposition} \label{POV}     
The most general translation covariant bounded POV measure $\tau$ is given by eqs.\  {\rm (\ref{Integral2})}, {\rm (\ref{Density})} and {\rm (\ref{PsiX3})},  where the quantities $K_{\gamma \sigma}(k)$ satisfy the condition {\rm (\ref{Condition})} and, if we impose the normalization condition {\rm (\ref{Normal1})}, also eq.\  {\rm (\ref{Normal2})}. We have 
\begin{equation}  
\tau(I) \psi = 0 \qquad {\rm if} \quad \psi \in {\cal H}''. 
\end{equation} 
\end{proposition}
Note that for $d = 1$ these results apply to a time-of-arrival observable.   

In an asymptotically complete \cite{StW, Jost} theory without massless particles, the subspace ${\cal H}''$ contains the vacuum and the one-particle states. It is physically clear that those states cannot individuate an event. The subspace ${\cal H}'$ contains scattering states, which can be described in terms of two or more incoming or outgoing particles. The description of $\cal H$ in terms of asymptotic states is physically interesting, because it deals with a situation in which the space-time position of an event, for instance a collision, is measured by means of operations performed in a far-away region, as it happens, for instance, in the famous conceptual position measurement by means of a microscope, discussed by Heisenberg \cite{Heisenberg} and reconsidered by Mead \cite{Mead} in the presence of the gravitational interaction. This point of view could also provide the starting point for the introduction of space-time concepts in a pure S-matrix theory.

\section{Poincar\'e covariant POV measures.}  

The covariance condition (\ref{Covariance}) with respect to the proper orthochronous Poincar\'e group introduces some new constraints. It is easy to show that the linear space $\cal D$ is invariant under the Poincar\'e transformations and if $\phi, \psi \in {\cal D}$ we have
\begin{equation}  
\rho(U(y, a) \psi, U(y, a) \phi, \Lambda(a) x + y) = \rho(\psi, \phi, x).
\end{equation} 
Since the translation covariance has already been exploited, it is sufficient to impose that
\begin{equation} 
\rho(U(0, a) \psi, U(0, a) \phi, 0) = \rho(\psi, \phi, 0).
\end{equation} 
It follows that $U(0, a)$ defines an unitary representation of $SL(2, C)$ in the quotient space ${\cal D}/{\cal D}_0$ and in its completion $\tilde{\cal H}$. We indicate this representation by $\tilde U(a)$. The operator $h$ is an intertwining operator, namely we have
\begin{equation} \label{Intertw}
\tilde U(a) h = h U(0, a).
\end{equation}   

In order to proceed, we have to examine the representations $U(x, a)$ and $\tilde U(a)$ with more detail.  We consider again the case $d = 4$. We remark that the subspace ${\cal H}'$ is invariant under the representation $U(x, a)$ and we indicate by $U'(x, a)$ the restriction of $U(x, a)$ to ${\cal H}'$.  We decompose this representation into a direct integral of irreducible unitary representations (IURs) of $\tilde {\cal P}$ \cite{Wigner}. Of course, only positive-energy representations appear in this decomposition. Since the four-momentum spectrum is absolutely continuous, we can disregard zero-mass representations and consider only positive-mass IURs, which are labelled by the mass $\mu$ and the spin $j$. 

The Hilbert space ${\cal H}'$ is decomposed into a direct integral of spaces in which IURs of $\tilde{\cal P}$ operate. A vector $\psi \in {\cal H}'$ is described by a wave function of the kind  $\psi_{\sigma j m}(k)$, where the index $\sigma$ labels the spaces in which equivalent IURs operate. For instance, in a two-particle state $\sigma$ describes the centre-of-mass helicities \cite{JW}. It is not necessary to specify the mass $\mu$, since it is a function of $k$. The norm is given by 
\begin{equation}  \label{Norm3}
\|\psi\|^2 = \int_V \sum_{\sigma j m} |\psi_{\sigma j m}(k)|^2  \, d^4 k,  
\end{equation}
where $V$ is the open future cone. 

For fixed $\sigma$, $\mu$ and $j$,  the group  $\tilde{\cal P}$  acts in the way described by Wigner \cite{Wigner}. We choose for each four-momentum $k \in V$  
an element $a_k \in SL(2, C)$ with the property 
\begin{equation}
k = \Lambda(a_k)q(\mu), \qquad  q(\mu) = (\mu, 0, 0, 0), 
\qquad \mu = (k_{\alpha} k^{\alpha})^{1/2}
\end{equation}
and we indicate by $R^j_{mm'}(u)$ the $(2j +1)$-dimensional IUR of $SU(2)$. Then we have
\begin{equation} \label{Poincare}
[U'(x, a)\psi]_{\sigma j m}(k) = \exp(ik_{\alpha} x^{\alpha}) R^j_{mm'}(u) \psi_{\sigma j m'}(k'), 
\end{equation}
where
\begin{equation} \label{KaP}
k' = \Lambda(a^{-1})k, \qquad  u = a_k^{-1} a a_{k'} \in SU(2).
\end{equation}  

In order to describe the representation $\tilde U$, we consider its direct integral decomposition into IURs of $SL(2, C)$.  Their matrix elements $D^{M c}_{jmj'm'}(a)$  are treated in refs.\ \cite{Naimark, GGV, Ruhl, ST}. There are two series of IURs: the principal series with $c$ imaginary and $M$ integral or half-integral, and the supplementary series with $-1 < c <1$ and $M = 0$. The representations  $D^{M c}$ and $D^{-M -c}$ are unitarily equivalent. One should not forget the trivial one-dimensional representation. Since the symbol $D^{0, \pm1}$ does not appear in the list given above, we use it to indicate the one-dimensional representation. It has only the matrix element $D^{0, \pm1}_{0000}(a) = 1$. This convention is partially justified by continuity arguments.

The restriction of these representations to the subgroup $SU(2)$ is given by
\begin{equation} 
D^{M c}_{jmj'm'}(u) = \delta_{jj'} R^j_{mm'}(u), \qquad u \in SU(2)
\end{equation}    
and the possible values of the indices $j, m$ are
\begin{displaymath}  
j = |M|, |M| + 1,\ldots, \qquad {\rm for} \qquad c \neq \pm 1,
\end{displaymath} 
\begin{displaymath}  
j = |M| = 0, \qquad {\rm for} \qquad c = \pm 1,
\end{displaymath} 
\begin{equation}  \label{Values}
m = -j, -j + 1,\ldots, j.
\end{equation}
In the following it is understood that all the quantities that depend on these indices vanish if the these relations are not satisfied. 

We consider the direct integral decomposition the Hilbert space $\tilde{\cal H}$  into irreducibles spaces labelled by the variable $\gamma$
\begin{equation} 
\tilde{\cal H} = \int^{\oplus}_{\Gamma} \tilde{\cal H}_{\gamma} \, d\omega(\gamma).
\end{equation}                    
The variable $\gamma$ stands for the parameters $c$ and $M$ that label the equivalence classes of IURs of $SL(2, C)$ and an index $\nu$ that distinguishes the spaces where equivalent IURs operate. $\Gamma$ is a set of points labelled by these parameters and $\omega$ is a positive measure on $\Gamma$. An element $\Psi \in \tilde{\cal H}$ can be described by the quantity $\Psi_{\gamma l n} = \Psi_{\nu c M l n}$. Its norm is given by
\begin{equation}  \label{Norm4}
\|\Psi\|^2 = \int_{\Gamma} \sum_{ln} |\Psi_{\gamma ln}|^2 \, d \omega(\gamma)
\end{equation}
and the representation $\tilde U$ acts in the following way
\begin{equation} 
[\tilde U(a) \Psi]_{\nu c M l n} = \sum_{l'n'} D^{M c}_{lnl'n'}(a) \Psi_{\nu c M l' n'}.
\end{equation} 
               
Now we have to adapt the formulas found in the preceding Section to the description given above of the spaces ${\cal H}'$ and $\tilde{\cal H}$. In the first case, we have just to replace the index $\sigma$ by the set of indices $\{\sigma, j, m\}$, as we have done in replacing eq.\ (\ref{Norm}) by eq.\ (\ref{Norm3}). In the second case, we have to replace the index $\gamma$ by the set of indices $\{\gamma, l, n\}$, where $\gamma = \{\nu, c, M\}$. Since $c$ is a continuous parameter, the sum over $\gamma$ has to be replaced by an integral with respect to the measure $d\omega(\gamma)$, which also implies a sum over the indices $\nu$ and $M$. In this way, for example, we pass from eq.\ (\ref{Norm2}) to eq.\ (\ref{Norm4}).

By means of these substitutions, the eq.\ (\ref{Psi}) takes the form
\begin{equation}  
[h \psi]_{\gamma l n} = \Psi_{\gamma l n} = (2 \pi)^{-2} \int_V \sum_{\sigma j m} K_{\gamma l n \sigma j m}(k) \psi_{\sigma j m}(k) \, d^4 k.
\end{equation} 
From the intertwining property (\ref{Intertw}) we obtain   
\begin{equation} \label{Intertw1} 
\sum_{l' n'} D^{M c}_{lnl'n'}(a)  K_{\nu c M l'n' \sigma j m}(k') = \sum_{m'} K_{\nu c M l n \sigma j m'}(k) R^j_{m' m}(u),
\end{equation}   
where $k'$ and $u$ are given by eq.\ (\ref{KaP}). 
If we put $a = a_k a_{k'}^{-1}$, we get $u = 1$ and, using the representation property,  
\begin{equation} 
\sum_{l' n'} D^{M c}_{lnl'n'}(a_{k'}^{-1})  K_{\nu c M l'n' \sigma j m}(k') =
\sum_{l' n'} D^{M c}_{lnl'n'}(a_k^{-1})  K_{\nu c M l'n' \sigma j m}(k). 
\end{equation} 
We see that this is a Lorentz invariant function of $k$, which depends only on $\mu$. Then we can write
\begin{equation} 
K_{\nu c M l n \sigma j m}(k) = \sum_{l' n'} D^{M c}_{lnl'n'}(a_k) F_{\nu c M l'n' \sigma j m}(\mu). 
\end{equation}   
If we substitute this formula into eq.\ (\ref{Intertw1}) we obtain 
\begin{equation}  
\sum_{n'} R^l_{nn'}(u)  F_{\nu c M l n' \sigma j m}(\mu) = \sum_{m'} F_{\nu c M l n \sigma j m'}(\mu) R^j_{m' m}(u)
\end{equation}
and from the Schur lemma we obtain  
\begin{equation}  
F_{\nu c M l n \sigma j m}(\mu) = F^j_{\nu c M \sigma }(\mu) \delta_{l j} \delta_{n m} 
\end{equation} 
and in conclusion
\begin{equation} 
K_{\nu c M l n \sigma j m}(k) = D^{M c}_{lnjm}(a_k) F^j_{\nu c M \sigma }(\mu). 
\end{equation}

By taking this formula into account and by adding the new representation indices, the eqs.\ (\ref{Density}), (\ref{PsiX3}), (\ref{Condition}) and (\ref{Normal2}) take the form              \begin{equation} \label{Density2}
\rho(\psi, x) =  \int_{\Gamma} \sum_{ln} |\Psi_{\gamma l n}(x)|^2 \, d \omega(\gamma),
\end{equation}                   
\begin{equation} \label{PsiX4}
\Psi_{\gamma ln}(x) = (2 \pi)^{-2}  \int_V \exp(-i x_{\alpha} k^{\alpha}) \sum_{\sigma j m} D^{M c}_{lnjm}(a_k) F^j_{\gamma \sigma}(\mu) \psi_{\sigma j m}(k) \, d^4 k,
\end{equation} 
\begin{equation} \label{Condition2}
\int_{\Gamma} \left|\sum_{\sigma} F^j_{\gamma \sigma}(\mu) c_{\sigma} \right|^2 \, d\omega(\gamma) \leq \sum_{\sigma} |c_{\sigma}|^2,
\end{equation}
\begin{equation}  \label{Normal3}
 \int_{\Gamma} \overline{F_{\gamma \sigma}^j(\mu)} F_{\gamma \sigma'}^j(\mu) \, d\omega(\gamma) = \delta_{\sigma \sigma'}.
\end{equation}    

In conclusion, we have
\begin{proposition}  \label{POV2}
The most general bounded Poincar\'e covariant POV measure $\tau$ on the Minkowski space-time is given by eqs.\ {\rm (\ref{Integral2})}, {\rm (\ref{Density2})} and {\rm (\ref{PsiX4})}, in terms of the measure $\omega$, and of the function $F_{\gamma \sigma}^j(\mu)$ satisfying the condition {\rm (\ref{Condition2})}. The normalization condition {\rm (\ref{Normal1})} is equivalent to eq.\ {\rm (\ref{Normal3})}.
\end{proposition} 

It is interesting to remark that the IURs of $SL(2, C)$ belonging to the supplementary series and the one-dimensional representation may appear in the decomposition of $\tilde U$ on which the construction of the POV measure is based. On the contrary, they do not appear in the direct integral decomposition, based on the Plancherel formula \cite{Naimark, GGV, Ruhl},  of the unitary representation $U'(0, a)$ which acts on the physical Hilbert space ${\cal H}'$.  This could not happen if the intertwining operator $h$, which is defined on $\cal D$ had a unitary extension to the whole Hilbert space $\cal H$. The existence of this extension, however, does not follow from our assumptions. 

If the theory is invariant under the dilatations
\begin{equation}  \label{Dilatation}
\psi_{\sigma j m}(k) \to \psi'_{\sigma j m}(k) = \lambda^2 \psi_{\sigma j m}(\lambda k),
\end{equation}     
the covariance under dilatations requires
\begin{equation} 
\rho(\psi', x) = \lambda^{-4} \rho(\psi, \lambda^{-1}x).
\end{equation}  
This condition is equivalent to the requirement that the functions $F_{\gamma \sigma}^j(\mu)$ do not depend on $\mu$.

\section{Baricentric events.}                                      

The general formulas given in the preceding Sections describe a very large class of  covariant POV measures.  Now we have to discuss how some physical requirements can be used to obtain more definite results. We have already discussed the normalization requirement (\ref{Normal1}). Another interesting condition is to require that the space coordinates $X^1, X^2, X^3$ coincide with the coordinates of the centre-of-mass at the time $X^0$. The formulation of this condition for a quantum system is somehow ambiguous and it is useful to discuss first the classical relativistic case. We indicate by 
\begin{equation} 
L^{\alpha \beta}(x) = L^{\alpha \beta} - x^{\alpha} P^{\beta} + x^{\beta} P^{\alpha}, \qquad 
L^{\alpha \beta} = L^{\alpha \beta}(0),
\end{equation} 
the relativistic angular momentum tensor with respect to the point $x \in {\cal M}$. The world-line of the centre of mass contains $x$ if we have \cite{Moller}
\begin{equation} 
L^{10}(x) = L^{20}(x) = L^{30}(x) = 0.
\end{equation} 
If in center-of-mass system there is a non vanishing angular momentum, the position of the center-of-mass depends on the velocity of the observer and it is useful to work in a frame in which $P^1 = P^2 = P^3 = 0$. Then, the square of the spatial distance of the centre-of-mass from the origin, which does not depend on time, is given by
\begin{equation} 
\Xi = \sum_{\alpha = 1}^3 (x^{\alpha})^2 = (P^0)^{-2} \sum_{\alpha} (L^{\alpha 0})^2 = (P_{\gamma} P^{\gamma})^{-2} L_{\alpha \beta} P^{\beta} P_{\gamma} L^{\gamma \alpha}.
\end{equation} 
We adopt the last expression, which is Lorentz invariant.                   

In a quantum theory we have to use the Hermitian operator corresponding to the quantity $\Xi$ (there is some problem of ordering). It is defined on the dense space $\cal D$ introduced in Section 2 and, since it is positive, it has a self-adjoint extension and we can consider a wide class of functions $f(\Xi)$. It is natural to interpret the  operator $\theta(\Xi - \eta^2)$, as the spectral projector on the states in which the world-line of the centre of mass has a distance from the origin larger that $\eta > 0$. If we introduce the Casimir operators of $\tilde{\cal P}$ 
\begin{equation} 
C_1 = P_{\alpha} P^{\alpha}, \qquad
C_2 = - S_{\alpha} S^{\alpha}, \qquad
S^{\alpha} = {1 \over 2} \epsilon^{\alpha \beta \gamma \delta} P_{\beta} L_{\gamma \delta}
\end{equation} 
and the Casimir operators of $SL(2, C)$
\begin{equation} 
C_3 = {1 \over 2} L_{\alpha \beta} L^{\alpha \beta}, \quad
C_4 = {1 \over 8} \epsilon^{\alpha \beta \gamma \delta} L_{\alpha \beta} L_{\gamma \delta},
\end{equation}
we have
\begin{equation} 
\Xi = C_1^{-2} C_2 - (C_1)^{-1} C_3.
\end{equation}
 
The Casimir operators have the properties  
\begin{equation}
[C_1 \psi]_{\sigma j m}(k) = \mu^2 \psi_{\sigma j m}(k),  \qquad
[C_2 \psi]_{\sigma j m}(k) = \mu^2 j(j + 1) \psi_{\sigma j m}(k),    
\end{equation} 
\begin{equation}
[C_3 \Psi]_{\gamma l n}(0) = (M^2 + c^2 -1)  \Psi_{\gamma l n}(0), \qquad
[C_4 \Psi]_{\gamma l n}(0) = i M c \Psi_{\gamma l n}(0).    
\end{equation} 
Note that, since $\psi \in {\cal D}$, $\Psi_{\gamma l n}(x)$ is a differentiable function of $x$. Then from eq.\ (\ref{PsiX4}) we have 
\begin{displaymath} 
[f(\Xi) \Psi]_{\gamma l n}(0) =  (2 \pi)^{-2} \cdot
\end{displaymath}  
\begin{equation} 
\cdot  \int_V  \sum_{\sigma j m} D^{M c}_{lnjm}(a_k) F^j_{\gamma \sigma}(\mu) \psi_{\sigma j m}(k) f(\mu^{-2} (j(j + 1) - M^2 - c^2 +1)) \, d^4 k.
\end{equation} 
                                                    
According to our interpretation, a POV measure $\tau$ is strictly baricentric if we have
\begin{equation} \label{Baricentric}
\rho(f(\Xi) \psi, 0) = 0, 
\end{equation}
whenever the function $f$ vanishes in a neighborhood of zero, namely $f(\Xi) \psi$ represents a state in which the  of the world-line of the centre-of-mass does not meet a neighborhood of the origin. It could seem natural to require that the density (\ref{Baricentric}) vanishes in the same neighborhood, but this is not permitted by Proposition \ref{Positivity}. The condition (\ref{Baricentric}) implies that  
\begin{equation}
F^j_{\nu M c \sigma}(\mu) \neq 0 \qquad {\rm only\,\, if} \qquad j(j + 1) - M^2 - c^2 +1 = 0, 
\end{equation} 
namely if   
\begin{equation}
j = M = 0, \qquad c = \pm 1.
\end{equation} 
We see that in the definition of $\tau$ only the trivial one-dimensional representation $D^{0 1}$ can appear. Moreover, $\tau$ must vanish on all the subspaces of ${\cal H}'$ which correspond to non vanishing values of the index $j$ and it cannot be normalized in the sense of eq.\ (\ref{Normal1}), which requires that eq.\ (\ref{Normal3}) is satisfied for all the values of $j$ and $\mu$.  

If we consider a system composed of two free spinless point particles, $j$ is the angular momentum in their centre-of-mass and the condition $j = 0$ means that the two particles meet as closely as it is permitted by the indeterminacy relations; otherwise, the event does not take place. This point of view is similar to the one discussed in ref.\ \cite{Leon}, which deals with the time-of-arrival of a single relativistic particle at a fixed point in three dimensions. However, one can also consider events which correspond, in the classical case, to the centre-of-mass at the time in which the distance between the two particles takes its minimum value. An event of this kind happens for arbitrary values of the angular momentum $j$. 

In order to define a normalized POV measure which is as baricentric as possible, we have to minimize, for every value of $j$, the expression $j(j + 1) - M^2 - c^2 +1$, namely to impose the condition
\begin{displaymath} 
F^j_{\nu M c \sigma}(\mu) \neq 0 \qquad {\rm only\,\, if} 
\end{displaymath}   
\begin{equation}  \label{NearBar}
M = j, \qquad c= 1 \quad {\rm for} \quad j = 0 \qquad {\rm and} \quad c = 0 \quad {\rm for} \quad j > 0. 
\end{equation}   
Note that in this case the measure $\omega(\gamma)$ is discrete and the corresponding integral can be replaced by a sum, namely we can write
\begin{displaymath} 
\rho(\psi, x) = (2 \pi)^{-4} \int_V \int_V \exp(i x_{\alpha} (k^{\prime \alpha} - k^{\alpha})) \sum_{\nu j \sigma' m' \sigma m} D^{j c}_{jm'jm}(a_{k'}^{-1} a_k) \cdot
\end{displaymath}
\begin{equation} \label{QuasiBar}
\cdot \overline{F^j_{\nu \sigma'}(\mu')} F^j_{\nu \sigma}(\mu) \overline{\psi_{\sigma' j m'}(k')} \psi_{\sigma j m}(k) \, d^4 k' \, d^4 k, \qquad F^j_{\nu \sigma}(\mu) = F^j_{\nu j c \sigma}(\mu).
\end{equation}  
Note that there is no interference term between states with different $j$. The POV measures which satisfy the condition (\ref{NearBar}) will be called {\it quasi-baricentric}. We shall discuss some of their properties in the next Sections.

\section{The coordinate operators.} 

Even if the coordinates of an event are described more completely by the POV measure $\tau$, it is interesting to consider the (non self-adjoint) operators $X^{\alpha}$ defined by eq.\ (\ref{Coordinates}). Note that they can be used to calculate the average values of the coordinates, but not, in general, to obtain in the usual way the details of their statistical distributions. As we see from eq.\ (\ref{Commu3}), these operators are more meaningful when the normalization condition (\ref{Normal}) is satisfied and we consider only this case. Then we have to disregard the Hilbert subspace ${\cal H}''$ and consider only states with an absolutely continuous four-momentum spectrum. 

From eqs.\ (\ref{Integral2}) and  (\ref{Density2}) we obtain
\begin{equation}   \label{Coordinates2}
(\psi, X^{\alpha} \psi) = \int x^{\alpha} \rho(\psi, x) \, d^4 x = \int x^{\alpha} \int_{\Gamma} \sum_{ln} |\Psi_{\gamma l n}(x)|^2 \, d \omega(\gamma) \, d^4 x.
\end{equation}         
If $\psi \in {\cal D}$, from eq.\ (\ref{PsiX4}) we have
\begin{displaymath} 
x_{\alpha} \Psi_{\gamma ln}(x) = -i (2 \pi)^{-2} \int_V \exp(-i x_{\alpha} k^{\alpha}) \cdot
\end{displaymath}
\begin{equation} 
\cdot \sum_{\sigma j m} \frac{\partial}{\partial k^{\alpha}} \left( D^{M c}_{lnjm}(a_k) F^j_{\gamma \sigma}(\mu) \psi_{\sigma j m}(k) \right) \, d^4 k.
\end{equation}                                                 
The derivative in the right hand side is composed of three terms, which, when substituted into eq.\ (\ref{Coordinates2}), give rise to three contributions 
\begin{equation}   
(\psi, X^{\alpha} \psi) =  A^{\alpha} + B^{\alpha} + C^{\alpha}.
\end{equation}
By taking into account the normalization condition (\ref{Normal3}) and the representation property, after some calculations we obtain     
\begin{displaymath}   
A_{\alpha} = \int_V \int_{\Gamma} \sum_{\sigma' j' m' \sigma j m} S^{M c}_{\alpha j' m' j m}(k) \cdot
\end{displaymath}   
\begin{equation}   
\cdot \overline{F_{\Gamma \sigma'}^{j'}(\mu)}  F_{\gamma \sigma}^j(\mu) \overline{\psi_{\sigma' j' m'}(k)}\psi_{\sigma j m}(k) \, d^4 k \,d \omega(\gamma), 
\end{equation}       
\begin{equation}   
B^{\alpha} = \mu^{-1} \int_V k^{\alpha}  \sum_{\sigma \sigma' j m} T^j_{\sigma' \sigma}(\mu)  \overline{\psi_{\sigma' j m}(k)} \psi_{\sigma j m}(k) \, d^4 k, 
\end{equation}
\begin{equation}   
C_{\alpha} =  -i \int_V \sum_{\sigma j m} \overline{\psi_{\sigma j m}(k)} \frac{\partial}{\partial k^{\alpha}} \psi_{\sigma j m}(k) \, d^4 k,
\end{equation}  
where we have introduced the Hermitiam matrices
\begin{equation} \label{V}  
S^{M c}_{\alpha j' m' j m}(k) = -i \left[\frac{\partial}{\partial k^{\alpha}} D^{M c}_{j' m' j m}(a_{k'}^{-1} a_k) \right]_{k' = k},
\end{equation} 
\begin{equation}   \label{Umu}
T^j_{\sigma' \sigma}(\mu) = - i \int_{\Gamma} \overline{F_{\gamma \sigma'}^j(\mu)} \frac{\partial}{\partial \mu} F_{\gamma \sigma}^j(\mu) \, d\omega(\gamma). 
\end{equation}
The term $C^{\alpha}$ has a familiar form and it is covariant under translations, but not under the Lorentz group. The other terms are translation invariant. The term $B^{\alpha}$ vanishes if $F_{\gamma \sigma}^j(\mu)$ does not depend on $\mu$, as it necessarily happens in dilatation invariant theories. 

In order to compute the quantities (\ref{V}), we use the following expression for the Wigner boosts:
\begin{equation}   
a_k = (2 \mu (\mu + k^0))^{- 1/2} (\mu + k^0 + k^s \sigma^s).
\end{equation}                                               
Here and in the following the indices $r, s, t$ take the values $1, 2, 3$.  If we put $q^{\alpha} = k^{\alpha}- k^{\prime\alpha}$ and we disregard quadratic and higher order terms in these differences, we have  
\begin{equation}   
a_{k'}^{-1} a_k = 1 - \frac{q^0 k^r \sigma^r}{2 \mu^2} +
\frac{q^r \sigma^r}{2 \mu} + \frac{k^s \sigma^s k^r q^r}{2 \mu^2 (\mu + k^0)} +  i \frac{\epsilon^{rst} q^r k^s \sigma^t}{2 \mu (\mu + k^0)} + \ldots
\end{equation} 
If the quantites $\theta^r$ and $\zeta^r$ are infinitesimal, we have  
\begin{displaymath}   
D^{M c}_{j' m' j m}(1 - \frac{i}{2} \theta^r \sigma^r + \frac{1}{2} \zeta^r \sigma^r + \ldots) =
\end{displaymath}
\begin{equation}
= \delta_{j'j} \delta_{m'm} -i \theta^r \delta_{j'j} M^{r j}_{m'm} -i \zeta^r N^{rMc}_{j'm'jm} + \ldots,
\end{equation} 
where $M^{r j}_{m'm}$ are the usual angular momentum matrices and  $N^{rMc}_{j'm'jm}$ are the generators of the $SL(2, C)$ boosts, which can be found (with different notations) in ref.\ \cite{Naimark}. From these formulas we obtain
\begin{equation} 
S^{M c}_{0 j' m' j m}(k) = \frac{k^r}{\mu^2} N^{rMc}_{j'm'jm},  
\end{equation}   
\begin{displaymath} 
S^{M c}_{r j' m' j m}(k) = - \frac{1}{\mu} N^{rMc}_{j'm'jm} -
\end{displaymath}
\begin{equation}
- \frac{k^r k^s}{\mu^2 (\mu + k^0)} N^{sMc}_{j'm'jm} + \frac{\epsilon^{rst} k^s}{\mu (\mu + k^0)} \delta_{j'j} M^{tj}_{m'm}.
\end{equation}

Now we consider with more detail quasi-baricentric events, namely we assume that the condition (\ref{NearBar}) is satisfied. Then we have only terms with $j' = j = M$ and the value of $c$ is fixed. From the formulas given in ref.\ \cite{Naimark} we have 
\begin{equation} 
N^{rMc}_{jm'jm} = 0, \qquad {\rm if} \qquad Mc = 0.
\end{equation}  
Then, by means of eq.\ (\ref{Normal3}), we obtain the simpler expressions
\begin{equation} 
A^0 = 0, \qquad 
A^r = - \int_V \frac{\epsilon^{rst} k^s}{\mu (\mu + k^0)}  \sum_{\sigma j m' m} \overline{\psi_{\sigma j m'}(k)} M^{t j}_{m'm} \psi_{\sigma j m}(k) \, d^4 k.
\end{equation} 
Note that the same expression can be obtained for different choices of the parameters which define the POV measure $\tau$.  For instance, in the term with $j' = j = M = 0$ we can choose an arbitrary value of $c$ without affecting the operators $X^{\alpha}$.

We want to show that if the POV measure is normalized and quasi-baricentric, the operators $X^{\alpha}$ can be written in a form similar to the one suggested in ref.\ \cite{JR}, namely
\begin{equation}  \label{HatX} 
X^{\alpha} = (P_{\gamma} P^{\gamma})^{-1} (P_{\beta} L^{\alpha \beta} - P^{\alpha} (D -2i)), 
\end{equation}
where $L^{\alpha \beta}$ are the components of the relativistic angular momentum operator and $D$ is given by
\begin{equation}   
D = i \left( k^{\alpha} \frac{\partial}{\partial k^{\alpha}} +2 \right) - \mu T(\mu),
\end{equation}
where $T(\mu)$ is the matrix defined by eq.\ (\ref{Umu}) which acts on the index $\sigma$ of the wave function.  In a theory with dilatation symmetry the term $T(\mu)$ vanishes and the operator $D$ has a self-adjoint extension which is the generator of the dilatations defined by eq.\ (\ref{Dilatation}). In the general case, $D$ is just an Hermitian operator defined on the domain $\cal D$. It describes the clock which is necessarily present in the object that defines the event and $T(\mu)$ can be interpreted as a kind of proper time delay. 

The operators $L^{\alpha \beta}$ are the generators of the Lorentz transformations defined by eq.\ (\ref{Poincare}). They are given by 
\begin{equation}  
L_{\alpha \beta} = i \left( k_{\alpha} \frac{\partial}{\partial k^{\beta}} - k_{\beta} \frac{\partial}{\partial k^{\alpha}} \right) + \hat L_{\alpha \beta},
\end{equation}                                                       
where $\hat L_{\alpha \beta}$ are matrices which represent infinitesimal rotations and act on the index $m$ of the wave function. They have the form
\begin{equation}   
\hat L^{rs} =  \epsilon^{rst} M^t, \qquad 
\hat L^{0r} = - \hat L^{r0} = - \frac{1}{\mu + k^0}  \epsilon^{rst} k^s M^t. 
\end{equation}                                                               
One can easily verify that eq.\ (\ref{HatX}) follows from these formulas.  

We have seen that eq.\ (\ref{HatX}) holds for all the normalized quasi-baricentric POV measures and it agrees with the formula given in ref.\ \cite{JR} if the theory is symmetric under dilatations. Since the event lies on the world-line of the center-of-mass, the dynamical aspects concern only the clock described by the operator $D$. The other aspects have simply a kinematical, namely group-theoretical, character.

\section{Coordinates as definite observables.}  
                                
In this last Section we discuss the conditions that permit the determination of the coordinates of an event, in suitably chosen states, with an arbitrary precision. Following ref.\ \cite{Giannitrapani1}, we introduce the definition:
\begin{definition}
We say that an observable described by a POV measure $\tau$ is ``definite'' (or simply that $\tau$ is definite) if, whenever the set $I$ has a non empty interior we have
\begin{equation} \label{Definite}
\|\tau(I)\| = 1.
\end{equation}  
\end{definition} 
This means that, with an apropriate choice of the state $\psi$, the probability that the result of the observable lies in $I$ can be made as near to $1$ as we want.

If the POV measure is translation covariant, it is sufficient to impose the condition (\ref{Definite}) for the sets $I$ which form a fundamental system of neighborhoods of the origin. Another equivalent condition is to require the existence of a sequence $\{\psi^{(\lambda)}\}$ of normalized vectors with the property
\begin{equation} 
\lim_{\lambda \to \infty} (\psi^{(\lambda)}, \tau(I) \psi^{(\lambda)}) = 1
\end{equation}  
when $I$ is an arbitrary neighborhood of the origin. An equivalent property is
\begin{equation} 
\lim_{\lambda \to \infty} \rho(\psi^{(\lambda)}, x) = \delta^d(x)
\end{equation} 
in the sense of distribution theory. By means of a Fourier transformation and of eqs.\ (\ref{Density}) and (\ref{PsiX3}), we can also write
\begin{equation} \label{Definite2}
\lim_{\lambda \to \infty} \int_V \sum_{\gamma \sigma \sigma'} \overline{K_{\gamma \sigma'}(k')}  K_{\gamma \sigma}(k'+ k) \overline{\psi^{(\lambda)}_{\sigma'}(k')} \psi^{(\lambda)}_{\sigma}(k'+ k) \, d^d k' =  1
\end{equation}
always in the sense of distribution theory.
                
A sufficient condition can be obtained by considering the particular choice
\begin{equation} 
\psi^{(\lambda)}_{\sigma}(k) = \lambda^{-d/2} c_{\sigma}(k) \phi(\lambda^{-1} k),
\end{equation}
where $\phi(k)$ is a test function with compact support $J \subset V$ and
\begin{equation} \label{NormPhi}
\int |\phi(k)|^2 \, d^d k  = 1,
\end{equation}  
\begin{equation} \label{NormC}
\sum_{\sigma} |c_{\sigma}(k)|^2 = 1.
\end{equation}
We are assuming that for $\lambda$ sufficiently large the set $\lambda J$ is contained in the region (possibly dependent on $\sigma$) where the wave function $\psi_{\sigma}(k)$ is defined.
  
Then the eq.\ (\ref{Definite2}), after a rescaling of the integration variable, takes the form   
\begin{equation}  
\lim_{\lambda \to \infty} \int r(\lambda k', \lambda k'+ k) \overline{\phi(k')} \phi(k'+ \lambda^{-1} k) \, d^d k' = 1,
\end{equation}         
where
\begin{equation}  
r(k', k) = \sum_{\gamma \sigma \sigma'} \overline{K_{\gamma \sigma',}(k')} K_{\gamma \sigma}(k) \overline{c_{\sigma'}(k')} c_{\sigma}(k) \leq 1.
\end{equation}

From these formulas, we find 
\begin{proposition}     
The translation covariant POV measure $\tau$ described in Proposition {\rm \ref{POV}} is definite if we can find the measurable functions $c_{\sigma}(k)$, which satisfy the normalization condition {\rm (\ref{NormC})}, in such a way that 
\begin{equation} \label{Definite3} 
\lim_{\lambda \to \infty} r(\lambda k', \lambda k'+ k) =1, \qquad k' \in J, \qquad k \in {\bf R}^d.
\end{equation} 
\end{proposition} 
This is not a necessary condition, but it shows that the definiteness property depends on the asymptotic behaviour of $K_{\gamma \sigma}(k)$.

Now we consider a quasi-baricentric Poincar\'e covariant POV measure of the kind described by  eq.\ (\ref{QuasiBar}).   Since there is no interference between terms with different $j$, we try to satisfy eq.\ (\ref{Definite2}) by means of wave functions which do not vanish only for a given pair of values of the indices $j, m$ and for these values of the indices are given by
\begin{equation} 
\psi^{(\lambda)}_{\sigma j m}(k) = \lambda^{-d/2} c_{\sigma}(\mu) \phi(\lambda^{-1} k),
\end{equation} 
with the normalization conditions (\ref{NormPhi}) and
\begin{equation} \label{NormC2}
\sum_{\sigma} |c_{\sigma}(\mu)|^2 = 1.
\end{equation}
Also in this case we obtain the condition (\ref{Definite3}) with
\begin{equation} 
r(k', k) = \sum_{\nu \sigma' \sigma} \overline{F^{j}_{\nu \sigma',}(\mu')} F^j_{\nu \sigma}(\mu) D^{j c}_{jmjm}(a_{k'}^{-1}a_k) \overline{c_{\sigma'}(\mu')} c_{\sigma}(\mu).
\end{equation}    

If we remark that $a_{\lambda k} = a_k$, we see that 
\begin{equation}  
\lim_{\lambda \to \infty} a_{\lambda k'}^{-1}a_{\lambda k'+ k} = 1,
\end{equation} 
it follows that
\begin{equation} 
\lim_{\lambda \to \infty} (r(\lambda k', \lambda k'+ k) - \hat r(\lambda k', \lambda k'+ k)) = 0, 
\end{equation} 
where   
\begin{equation} \label{Er2} 
\hat r(k', k) = \hat r(\mu', \mu) = 
\sum_{\nu \sigma' \sigma} \overline{F^{j}_{\nu \sigma'}(\mu')} F^j_{\nu \sigma}(\mu) \overline{c_{\sigma'}(\mu')} c_{\sigma}(\mu).
\end{equation} 

In conclusion, we have the following sufficient condition:
\begin{proposition}     
The Poincar\'e covariant and quasi-baricentric POV measure $\tau$ described by eq.\ {\rm (\ref{QuasiBar})} is definite if we can find the measurable functions $c_{\sigma}(\mu)$, which satisfy the normalization condition {\rm (\ref{NormC2})}, in such a way that, for some value of the index $j$, the expression defined by eq.\ {\rm (\ref{Er2})} has the property 
\begin{equation} 
\lim_{\mu \to +\infty} \hat r(\mu, \mu + c) =1
\end{equation} 
uniformly for $c$  belonging to any bounded interval of the real line.
\end{proposition}

\bigskip  \bigskip

\noindent {\it Acknowledgments:} I am grateful to Dr.\ R. Giannitrapani for many useful discussions.

\end{document}